\documentclass[conference]{IEEEtran}
\usepackage{blindtext, graphicx}
\let\labelindent\relax
\usepackage{enumitem, amsmath}

\ifCLASSINFOpdf
\else
\fi
\begin{document}
%
\title{Schema Matching using Machine Learning}




%
\author{\IEEEauthorblockN{Tanvi Sahay\IEEEauthorrefmark{1},
Ankita Mehta\IEEEauthorrefmark{1},
Shruti Jadon\IEEEauthorrefmark{1}}
\IEEEauthorblockA{\IEEEauthorrefmark{1} College of Information and Computer Sciences
\\
University of Massachusetts,
Amherst, MA 01002\\tsahay@cs.umass.edu, amehta@cs.umass.edu, sjadon@cs.umass.edu}}


\maketitle

\begin{abstract}
Schema Matching is a method of finding attributes that are either similar to each other linguistically or represent the same information. In this project, we take a hybrid approach at solving this problem by making use of both the provided data and the schema name to perform one to one schema matching and introduce creation of a global dictionary to achieve one to many schema matching. We experiment with two methods of one to one matching and compare both based on their F-scores, precision and recall. We also compare our method with the ones previously suggested and highlight differences between them.  
\end{abstract}


%
\IEEEpeerreviewmaketitle

\section{Introduction}
The schema of a database is the skeleton that represents its logical view. In other words, a schema describes the data contained in a database, with the name of each attribute in a relation and its data type contained in the relation's schema. Any time the different tables maintained by a peer management system need to be linked to each other or when one branch of a company is closed down and all its data needs to be redistributed to the database maintained by other branches or when one company takes over another company and all data of the child comapny needs to be integrated with that of the parent company, the need to match schemas of multiple relations with each other arises. In basic terms, schema matching can be explained as follows: Given two databases
$X(x_1, x_2, x_3 )$ and $Y(y_1, y_2, y_3 )$ with $x_n$ and $y_n$ representing
their attributes resepectively, we match a schema attribute to another either if it is linguistically similar(has a similar name) or if it represents the same data. Consider the Tables \ref{students} and \ref{grad-students}. Here, the ideal schema mappings would be: \textit{FName $+$ LName $=$ Name}, \textit{Major $=$ Maj$\_$Stream} and \textit{Address $=$ House No $+$ St Name}.  

\begin{table}[h]
\centering
\caption{Students}
\begin{tabular}{|c|c|c|c|c|}
\hline
FName & LName & SSN & Major & Address\\
\hline \hline
Shruti & Jadon & 123-aaa-aaaa & Computer Science & 1xx Brit Mnr\\
Ankita & Mehta & 234-bbb-bbbb & Mathematics & 2xx Boulders\\
Tanvi & Sahay & 456-ccc-cccc & Political Science & 3xx N Pleasant St\\
\hline
\end{tabular}
\label{students}
\end{table}

\begin{table}[h]
\centering
\caption{Grad-Students}
\begin{tabular}{|c|c|c|c|c|}
\hline
Name & ID & Maj\_Stream & House No & St name\\
\hline \hline
Shruti Jadon & 123aaa & CompSci & 1xx & Brit Mnr\\
Ankita Mehta & 23bbb4 & Math and Stats & 2xx & Boulders\\
Tanvi Sahay & 45cccc & PoliSci & 3xx & N Pleasant St\\
\hline
\end{tabular}
\label{grad-students}
\end{table}

Over the years, researchers have faced several issues when trying to automate the process of matching schemas of different relations. Because the schemas are created by human developers and are pertinent to a particular domain, human intervention is often required at one or multiple stages of the process to ensure proper schema matching, which makes this task quite labor intensive. The aim of automated schema matching is to reduce the involvement of a domain expert in the process to a minimum. Majorly, schema matching can be divided into two parts - one to one matching, where one attribute of table 1 matches with only one attribute of table 2 and one to many matching, where one attribute of table 1 may map to a combination of several attributes of table 2. In the above tables, (`Major',`Maj\_Stream') is a one-to-one matching and (`Address',[`House No',`St Name']) is a one to many matching. While one to one matching has been successfully automated using sophisticated machine learning techniques as well as by exploiting the schema structure, performing one to many schema matching still requires some form of human intervention. In general, matching can be done by taking into account either the data contained in the relations or the name of the attributes or both.

In this project, we explore two methods of performing one to one matching and suggest a new method of one to many mapping which is different from the ones that have been employed before. For one to one matching, we consider two appoaches, both based on utilizing a set of features to limit the set of candidate matches by clustering similar attributes together. In the first method, called centroid method, we cluster similar values of one table together into groups and compare each attribute of the second table with each cluster, to find the cluster that best matches with it. In the second method, called the combined method, we combine attributes of both tables into a single list and cluster all of them together to form groups containing similar fields from both tables. The centroid method, as we will see in the future sections, ensures that every attribute in the second table matches with at least one attribute in the first table. The combined approach on the other hand still has the possibility of an attribute in one table not matching with any other attribute in the second table. Each method will be discussed in more detail in the future sections and their tradeoffs as well as their performance with existing techniques will be compared as well. In addition to these techniques, we will discuss a new way of taking care of one to many matchings with minimum requirement of an external expert.

\section{Previous Work} 

\subsection*{\textbf{Database Schema Matching Using Machine Learning with Feature Selection}\cite{ref2}}
This paper is discussing about a tool called Automatch for automating the schema matching process. This approach consists of a global dictionary which is created by using schema examples and tuned by domain experts. Dictionary includes various clusters of attributes say R1, R2, R3 etc. It compares attributes of one schema (S1, S2, S3etc) with each of the dictionary attributes (R1, R2, R3 etc) and assign a weight based on probability formula of symmetry. The same is repeated with another schema and a path from schema 1 to schema 2 via the dictionary is chosen. The Minimum Weight Path determines which attribute of schema 1 is closely aligned with which schema 2 attribute.

While this method improves on its predecessors by including one-to-one attribute matching rather than just matching one attribute with a set of possible attributes, it still has the same problem that it does not consider the possibility of one attribute matching to a set of attributes.

\subsection*{\textbf{Semantic Integration in Heterogenous Databases using Neural Networks}\cite{ref1}}
This paper implemented schema matching using Machine Learning approach. It extracts the features of each column by using only their data values and these features, represented as vectors with each value lying in the range (0,1) are used as identifiers for that column. Then they are clustered together using a self-organizing map and their cluster centres are calculated. Using these cluster centres single hidden layer neural network with M outputs neurons (M = number of clusters) is trained and then tested with output as the similarity percentage of the attribute with each cluster.

While this method, known as SemaInt, provides the user with a similarity mapping of each attribute in one schema with a set of attributes in another, it does not take into account the fact one might map to a set of others as well.

\subsection*{\textbf{Corpus-based Schema Matching}\cite{ref4}}
This paper makes use of a corpora of schemas to prepare models of each attribute to be matched by making use of information provided by other attributes in the corpora similar to the ones being matched. Similar attributes are found by making use of learners such as name learner, text learner, context learner etc. and for matching attributes across two schemas, similarity of an attribute matching with the other based on the new `augmented' models is calculated. 

This method only considers one to one matching of attributes and cannot handle complex mappings like one to many or many to one. It also requires a significant amount of corpora to successfully learn good attribute models.

\subsection*{\textbf{Generic Schema Matching with Cupid}\cite{ref3}}
This paper explores a technique of matching which is schema based and not instance based. In the proposed method, heirarchical schemas are represented as trees and non-heirarchical schemas are generalied as graphs. Two types of matching scores, based on linguistic similarity i.e. similarity between schema attribute names, data types and domain etc. and based on structural similarity i.e. similarity based on context and vicinity are calculated and their average is assigned as the final matching score for a pair of attributes.

This method maintains a thesaurus for finding linguistic similarity and also makes use of information other than just the schema name, such as schema structure and relation of attributes with each when assigning scores.

\subsection*{\textbf{iMAP:Discovering Complex Semantic Matches between Database Schemas}\cite{ref5}}
iMAP introduces a new method of semi-automatically performing both one to one and one to many schema matching by converting the matching problem to a search problem in a relatively large search space of all possible schema mappings. For efficient searching, the paper proposes to make use of custom searchers based on concatenation of text, arithmetic operations over numeric attributes etc. and scoring each match to find the best possible matchings. Since the searchers are customized over type of data, they only search through a subset of search space, thus reducing system complexity. 

While this method achieves one to many mapping, it still requires a domain expert for creating custom searchers specific to a particular type of database. The method also makes use of only the data contained in the tables and not the schema names themselves.\\

\noindent
As we have seen, the methods shown above either focus only on one to one mappings or, when considering one to many mappings, do not take the actual schema names into account. One to one schema matching techniques also require a large amount of data to successfully train the machine learning models being employed, which may not necessarily be available. Our method presents a different approach in that we consider both one to one and one to many mapping and make use of both the data represented by the schema and the schema names themselves. The technique is not data intensive and requires minimum human intervention, requiring a domain expert only for the task of creating the one to many mapping dictionary.

\section{Methodology}
For the purpose of implementation, we have divided our task into two separate sections: One to Many Mapping and One to One Mapping. In all discussions that follow, we assume that we have a source schema S and a test schema T and our task is to map attributes present in the test schema to attributes present in the source schema. 

\subsection{Schema Data}
For this project, we perform all experiments on a subset of the medicare.gov data. We take two tables from the database, each of which represents the Inpatient Psychiatric Facility Quality Reports (IPFQR) of hospitals in the United States. One of the tables considers each hospital in the US and has a total of 85 attributes and 1644 data tuples with the field ``Provider$\_$number" taken as the primary key. The second table is the same data provided for only the best hospitals in each state, with State as the primary key. It has 74 attributes and 52 data tuples, one for each state and one for Washington DC and Puerto Rico each. Subsets of the two tables have been presented as Table \ref{HP-main} and Table \ref{SP-main} respectively. For all experiments, we have taken the first table i.e. the table containing all hospitals to be the source schema S and the second table to be the test schema T.

\begin{table}[h]
\centering
\caption{IPFQR data - General}
\begin{tabular}{|c|c|c|c|}
\hline
tr\_provider\_number & tr\_state & tr\_hbips\_2\_overall\_num & tr\_hie\_response\\
\hline \hline
10101 & AL & 23.7 & Yes\\
40014 & AR & 1.47 & No\\
34023 & AZ & 0.68 & Yes\\
\hline
\end{tabular}
\label{HP-main}
\end{table}

\begin{table}[h]
\centering
\caption{IPFQR data - Statewise}
\begin{tabular}{|c|c|c|c|}
\hline
ts\_state & ts\_s\_hbips\_2\_overall\_num & ts\_s\_hie\_yes\_count & ts\_start\_date\\
\hline \hline
AL & 2891.1 & 17 & 01/01/2015\\
AR & 844.77 & 10 & 01/01/2015\\
AZ & 4981.36 & 14 & 01/01/2015\\
\hline
\end{tabular}
\label{SP-main}
\end{table}

As can be seen from this small subset, the attributes are all domain centric and do not convey any semantic information about what the data contains, which is why any methods that match attributes semantically cannot be applied. 

Both data tables are stored in a single database and postgres combined with python has been used to access the data. Before performing schema matching, the attributes are cleaned up to allow uniformity across the schemas. Symbols such as $\%$ occurring in the schema names are converted to their actual name i.e. \textit{percent}. Certain integer or float type columns have char values such as `Not Available' which are converted to 0 and symbols such as $\%$ occurring in the data are removed as well. While storing this data in the database, schemas are normalized by converting all names to lower case and appending \textit{`tr\_'} to the source schema attributes and \textit{`ts\_'} to the test schema attributes. This is done to provide the user with a clear demarcation of which schema attributes belong to which table.

\subsection{One To Many Schema Matching}
One to many matching is done when a single attribute in the source schema matches with two or more attributes in the test schema. For example, the address of an individual can either be represented as `Address' in a sigle column or be broken down into `House No' and `Street no/name' as two independent columns. For achieving one to many schema matching, we propose the creation of a global dictionary that contains all possible mappings of a single attribute to multiple attributes and use this as a checkpoint to find out possible one to many mappings. This dictionary is represented as a set of key-value pairs, where keys are those attributes that can be broken down into several smaller ones and values are the corresponding set of attributes that together match with the key. In the example given above, `Address' will be considered a key and `House No' and `Street no/name' will be its corresponding values. When a key is present in the source schema and the key's corresponding values are present in the test schema, that set of attributes is separated as a one to many schema mapping. Attributes in S and T that have already been considered as a part of any one to many match will not be considered when looking at the one to one schema matching. 

The global dictionary created by us consists of the following key, value pairs:

\noindent
\textit{Key}: Address, Location, Addr, Loc, Residence\\
\textit{Value}: Street Name, S\_Name, St\_Name, Str\_Name, Stree\_Name, StName, St\_No, ST\_Number, Street\_No, S\_No, S\_Number, Street Number, StNumber, StNo, Apt\_Num, Apartment\_Number, Apartment Number, Apartment No, Apt\_Number, Apt\_No

\noindent
\textit{Key}: Name, PatientName\\
\textit{Value}: First Name, First\_Name, FName, F\_Name, Last\_Name, Last Name, LName, L\_Name\\

As can be seen, the keys and values consist of all possible ways of representing a particular attribute in order to capture a wider range of mappings. While the dictionary only consists of two possible one to many mappings at present, it can be extended with time by including more instances of such mappings. 

\subsection{One To One Schema Matching}
Once all one to many maps have been determined, we perform one to one matching on the remaining attributes. This is done by extracting meaningful descriptive features of each attribute and using these features to find similar attributes across the two schemas in consideration. After extracting features of each attribute, we experiment with two methods, namely: Centroid Method and Combined Method. Each method is evaluated using a measure called F1-Score, which is a combination of both precision and recall. Each section of one to one schema matching has been explained below.

\subsubsection*{\textbf{Feature Engineering and Feature Extraction}}

Features of an attribute are nothing but characteristics that decsribe the attribute in sufficient detail for it to be compared to and discriminated against other attributes and provide some idea of the similarity or dissimilarity between the compared attributes. These characteristics could either be based on the kind of data that attribute holds or based on schema information and specifications. Based on kind of data, these ``discriminators" can be data type, domain and range of data contained by the attribute, length of used space etc. and based on schema specifications, they can contain information about whether the attribute is a key or not and so on. Some of these features are binary, with values as either 0 or 1 while the others lie in the range [0,1] after normalization. Representing each attribute as this set of real values has several advantages. First, it allows us to perform mathematical operations that cannot be generally performed on text, which makes similarity computation easier. Second, by creating features manually, we can decide which aspects of an attribute to focus on when finding similar attributes. We have adapted from the feature set provided by \cite{ref1} to include a total of 20 ``discriminators". This set of features is not exhaustive and based on need, database type and type of information available, more features can be added to the list.

\textit{Features based on Schema Specification} - Based on schema information and specifications available, we create the following features: Type of data (Float, Int, Char, Boolean, date, time), length of fields provided by the user, whether the attribute is a key or not, whether the attribute has a unique condition or not, whether the attribute has a Not Null condition attached to it or not. This information can be easily extracted from the schema specified by the user. The UNIQUE constraint is considered only when it is specifically given i.e. when an attribute is not a key but is unique. For representation as a feature, we convert every value to a real number. For type of data, if the attribute is a float type, it is represented as 0 while if it is a time type, it is represented as 5. Length of the fields is already a real number and the remaining three attributes are provided a binary representation based on whether the attribute has a certain property or not (is a key - 1, not a key - 0 etc.).

\textit{Features based on Data Fields} - A lot of discriminatory features can be extracted based on the data contained within a particular column. We divide the data into three types - containing only numerics and provided as INT or FLOAT (age, salary), containing only char and provided as CHAR (name, city) and containing both and provided as VARCHAR (address). We create features that help discriminate between the three categories. Since there are certain types of information that only be contained in numeric attributes and certain that are specific to VARCHAR attributes, we divide our features into the following categories. 

\begin{enumerate}
\item Features for Numeric Data - The features specific to numeric data are:
\begin{itemize}[leftmargin=*]
\item Average - Average of all the entries in a specific column
\item Variance - Variance of a specific column. Variance is nothing but a measure of how far a set of values are spread from their mean.
\item Coefficient of variance - Coefficient of variance is another measure of variability but it aims to describ the dispersion in a way that does not depend on the variable's unit of measurement.
\item Minimum - The minimum value of a particular column
\item Maximum - The maximum value of a particular column
\end{itemize}
\item Features for Character Data - The features specific to character data are:
\begin{itemize}
\item Ratio of whitespace to length - Number of white space characters as opposed to text characters
\item Ratio of special characters to length - Number of special characters (`-', `\_', `(', `)', `$\setminus$' and so on) as opposed to text characters
\item Ratio of numeric to length - Number of numeric characters as opposed to total number of characters
\item Ratio of char to length - Number of plain text characters as opposed to total number of characters
\item Ratio of backslash to length - Number of backslash characters as opposed to total number of characters
\item Ratio of brackets to length - Number of brackets as opposed to total number of characters
\item Ratio of hyphens to length -  Number of hyphens as opposed to total number of characters
\end{itemize}
\item Features common to both - Certain features are common to both numeric and varchar attributes.
\begin{itemize}
\item Average Used Length - Length of attribute used as opposed to total specified length
\item Variance of Used length - Variance of the length used by values in a particular column
\item Coefficient of variance of used length - Coefficient of variance of the length used by values in a particular column 
\end{itemize} 
\end{enumerate}

For every attribute in both source and test schemas, a vector of length 20 is prepared that contains values for each of the 20 features decsribed above. To understand this with an example, consider the character attribute 

\noindent
\begin{align*}
ts\_state (CHAR(2)\ \ PRIMARY\ KEY)  
\end{align*}

\noindent
from the test table as shown in table \ref{SP-main}. Its feature vector has been shown in table \ref{features}. These are features before normalization and are hence not constrained between 0 and 1.

\begin{table}[h]
\centering
\caption{Features of attribute `ts\_state'}
\begin{tabular}{|c|c|c|c|}
\hline
Type of Data & 2 & Coeff of variance & 0.0\\
Length & 2 & Minimum & 0.0\\
Key & 1 & Maximum & 0.0\\
Unique & 0 & No of Whitespace & 0.0\\
Not Null & 1 & No of Special Char & 0.0\\
Average Length Used & 1.0 & Ratio of numerics & 0.0\\
Variance of Length & 0.0 & Ratio of chars & 1.0\\	
Var Coeff of Length & 0.0 & No of backslach & 0.0\\
Average & 0.0 & No of brackets & 0.0\\
Variance & 0.0 & No of hyphens & 0.0\\
\hline
\end{tabular}
\label{features}
\end{table}

Once features of each attribute have been extracted, we cluster similar attributes together using the methods explained in the following sections.

\subsubsection*{\textbf{Clustering and Linguistic Matching - Centroid Method}}

In the first approach, we cluster similar attributes of source schema together, essentially constraining the match search space for every test schema attribute.  For grouping alike attributes, we experiment with two different clustering methods: Kohonen Self Organising map and K-Means Clustering. Both have been explained as follows:

\textit{Self Organising Map\cite{SOM}: }A Kohonen Self-Organizing Map(SOM) is a type of Artificial Neural Network which is trained using unsupervised learning method in such a way that similar patterns in the input data are clustered together. A general architecture of the map has been shown in Figure 1, with the input layer having N nodes and the output layer having M nodes. Each output neuron is connected with every neuron in the input layer and each connection has a weight associated with it. For each input feature, a single output neuron is fired such that the weight vector associated with this neuron is closest to that input vector. Weights of all neurons near this activated neuron, including its own, are updated in such a way that it brings them closer to the input feature vector. Over several iterations, these weights are learned by the network and for any new input, the neuron with the weight vector closest to it is chosen as its class. For clustering, the input provided to SOM is a size 20 feature vector with number of examples equal to number of attributes in the source schema. Maximum clusters possible are equal to the number of output neurons. They can either be decided manually or chosen programatically. 

\begin{figure}[h]
\centering
\includegraphics[scale=0.4]{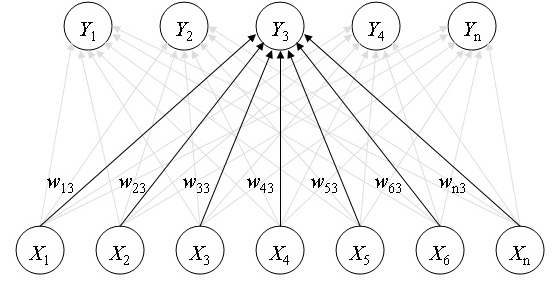}
\end{figure}

\textit{K-Means Clustering: }K-means is an iterative optimization algorithm for clustering that alternates between two steps. It randomly initializes k cluster centroids in a vector space D, with N data cases. From there on it repeats the following two steps until convergence: a) For each data point $x_i$ in X, find the nearest
centroid (cluster) to it (using Euclidean distance). Assign this data point to that cluster. b) For each of the clusters formed in step a), find the new centroids for each them, by finding out the mean in each of the clusters. Repeat step a) again. For our case, N data points, each of vector size equal to 20 and N equal to number of attributes in source schema are provided as input to the KMeans algorithm. 

For both clustering schemes, 7 cluster sizes are chosen and the best out of these is chosen by comparing their silhouette scores. Silhouette scoring is a measure of how similar an object is to its own cluster as opposed to other clusters. The method gives each clustering a score between -1 and 1, with values closer to 1 representing better clustering.

Once the clusters have been prepared, a centroid of each cluster is calculated. Centroid of a cluster is the average of all the feature vectors present in that cluster. To determine which set of source attributes a test attribute should be compared with, the euclidean distance of each test attribute with every cluster centroid is found. A test attribute $t_s$, only attempts to find one-to-one matches in the cluster whose centroid is closest to it and only that cluster of source attributes is considered as the candidate match space for $t_s$. Once a test attribute is assigned a search space, every candidate match in it is mapped to the test attribute linguistically and a similarity score is provided. 

Linguistic matching simply means matching the actual names of the attributes with each other. We have done this using edit distance which is a way of quantifying how dissimilar two strings (e.g., words) are to one another by counting the minimum number of operations required to transform one string into the other. Each source-test attribute pair is provided a probability of matching based on how close the two strings (names of the attributes) are.

\subsubsection*{\textbf{Clustering and Linguistic Matching - Combined Method}}
In this second approach, instead of only clustering similar attributes in the source schema and comparing centroids with the test attribute, we combine all attributes from both the schemas together and then apply clustering. Thus, input to the clustering method is a size 20 feature vector with number of examples equal to number of fields in source schema plus number of fields in test schema. For every cluster, we separate the source and test attributes and then for every test attribute in that cluster, we calculate its edit distance with every source attribute present in that cluster. Linguistic Matching cannot be applied to clusters having only train or test attributes because it shows that there is no mapping available for them. 

\subsubsection*{\textbf{Evaluation}}
For the purpose of assessment, we use F1-score to evaluate One to one Mappings obtained from Centroid and Combined method. F1 Score is an intrinsic cluster evaluation method to measure matching accuracy, which is the weighted average of precision and recall. Here, precision is the ratio of number of true positive results to the number of returned positive results and recall is the ratio of number of true positive results to the number of actual positive results.

To calculate precision and recall, manual mappings are computed first. Once we have the manual mappings, we compute the values of true positives, which are the number of correct mappings returned by our method, false positives, which are the number of values that were returned as a true match but should not have been and false negatives, which are the number of values that were not returned as a match but should have been. Based on these values, F1 score is defined as: 
\begin{equation}
F1 = \frac{2 * Precision * Recall}{Precision + Recall}
\end{equation}

\noindent
Where
\begin{equation}
Precision = \frac{True Positive}{True Positive + False Positive}
\end{equation}

\begin{equation}
Recall = \frac{True Positive}{True Positive + False Negative}
\end{equation}
    
\section{Experimentation and Results}
We perform all experiments on source and test schemas defined in the previous section. Multiple methods have been explored at each step. For Features, different set of features have been chosen and extended with time. The features are dataset dependent and for a different dataset, a different set of features may be extracted. For our particular dataset, we observe that inclusion of special character statistics plays an important role in refining the obtained clusters. Thus, after choosing the standard set of features explained in \cite{ref1}, we add features pertaining to our dataset. This improves our clusters, as confirmed via manual observation. 

We begin by performing one to many mapping on our dataset. There is only one such mapping present in our database, which is successfully identified by using the dictionary created by us, as explained in the previous section. Then, we move on to experiments in one-to-one mapping.

\begin{figure}[h]
\centering
\includegraphics[scale=0.45]{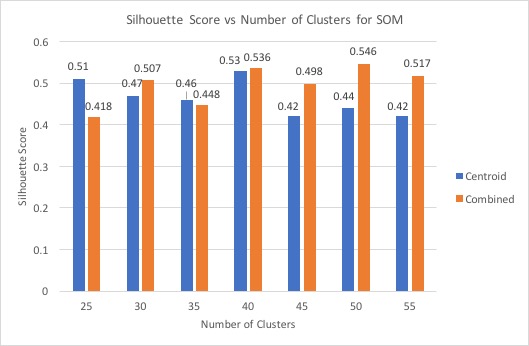}
\caption{Silhouette scores versus number of clusters for SOM}
\label{SOM_cluster}
\end{figure}

\begin{figure}[h]
\includegraphics[scale=0.35]{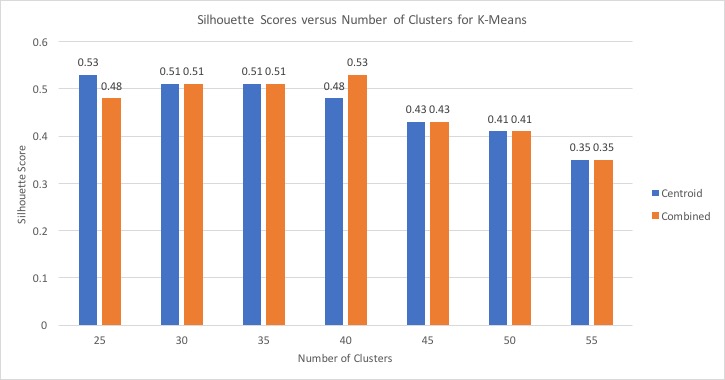}
\caption{Silhouette scores versus number of clusters for K-Means}
\label{KMeans_cluster}
\end{figure}

The first experiment we perform is with the clustering methods. For each of the two techniques, centroid and combined, we test SOM and K-Means clustering over a range of 7 clusters:- 25, 30, 35, 40, 45, 50 and 55. Ideally, clusters can be decided by the user depending on how narrow or wide they want their mappings to be. We choose these 7 values to cover a middle range of clusters, which are neither too narrow nor too wide. Figures \ref{SOM_cluster}, \ref{KMeans_cluster} and show the values of silhouette scores for each cluster size for both centroid and combined techniques. 

Based on the results shown above, we can see that SOM performs slightly better than K-Means with the highest silhouette score for a cluster size of 40for the centroid method and cluster size 50 for the combined method. We choose these as our optimal number of clusters for the two methods respectively and choose SOM as the clustering technique.

Next, to perform intra-cluster one to one matching between test and source attributes, we use two more methods in addition to edit distance. Euclidean Distance between the feature vectors of a test-source pair can be mathematically represented as:

\begin{equation}
\sqrt{(s_1-t_1)^2 + (s_2-t_2)^2 + ... + (s_n-t_n)^2}
\end{equation}

Here $s_i$ is the $i^{th}$ feature of a source attribute and $t_i$ is the $i^{th}$ feature of a test attribute. Cosine Similarity between two vectors is a measure of the cosine angle between them. For source and test vectors represented as $s$ and $t$ respectively, cosine similarity between them can be given as: 

\begin{equation}
\frac{\sum_{i=1}^n{s_it_i}}{\sqrt{\sum_{i=1}^ns_i^2}\sqrt{\sum_{i=1}^nt_i^2}}
\end{equation}

For each of the two methods i.e. centroid and combined clustering methods, F1-scores are calculated using all three distance measures for cluster size 40 for centroid method and 50 for combined method. The results have been compiled in tables \ref{results1} and \ref{results2} for centroid and combined methods respectively.

\begin{table}[h]
\centering
\caption{F1-scores for centroid method}
\begin{tabular}{|c|c|c|c|}
\hline
Distance Measure & Precision & Recall & F1 score\\
\hline \hline
\textbf{Edit Distance} & \textbf{0.527} & \textbf{1.0} & \textbf{0.690}\\
Euclidean Distance & 0.222 & 1.0 & 0.363\\
Cosine Similarity & 0.0555 & 1.0 & 0.105\\
\hline
\end{tabular}
\label{results1}
\end{table}

\begin{table}[h]
\centering
\caption{F1-scores for combined method}
\begin{tabular}{|c|c|c|c|}
\hline
Distance Measure & Precision & Recall & F1 score\\
\hline\hline
\textbf{Edit Distance} & \textbf{0.54} & \textbf{1.0} & \textbf{0.71}\\
Euclidean Distance & 0.194 & 1.0 & 0.325\\
Cosine Similarity & 0.027 & 1.0 & 0.054\\
\hline
\end{tabular}
\label{results2}
\end{table}

From the results above, we see that edit distance is a much better method of performing strict one-to-one mapping between source and test schema attributes. We also see the combined method performs slightly better than the centroid method, with the highest F1-score going up to 0.71. 

Comparing with results of \cite{ref4}, we see that our method is not as great in terms of performance as the corpus based method, which has the highest average F1-score of 0.87. \cite{ref1} provides its results in terms of output match similarity and reports a maximum similarity of 0.995 on the IBM AS/400 database. Using our method, we achieve a maximum similarity of 0.984 on the IPFQR dataset using edit distance and the centroid method. While our results are not at par with the currently available methods, we are still able to achieve good enough one-to-one mappings with a relatively small amount of data.

\section{Conclusion}
In this project, we tackle the problem of one-to-many and one-to-one schema matching using a small amount of data. We propose the creation of a global dictionary with all possible one to many mappings which may be extended over time to tackle the one-to-many matching case. This method would be particularly helpful for a closed domain problem, where only limited one-to-many mappings are possible. It also allows user customizability, so everytime a user adds a new attribute with a name that may occurr in a one-to-many mapping, they can add it to the dictionray and other users may take advantage of it. For one-to-one mappings, we base our methodology on creating an unsupervised clustering model for attributes in the source schema in order to limit the test attribute's search for matching attributes. We make use of both the data provided in the tables and the schema names themselves to find a strict one-to-one matching between each test and train attribute. 

Out of the two methods we propose, combined method performs better than centroid method for our particular dataset. We chalk this up to the fact that the combined method allow room for a test attribute to not match with any train attribute, which the centroid method does not. We also see that edit distance captures similarity between name strings much better than euclidean or cosine distance. This happens because our dataset has similar attributes having similar names. For a case where attribute representing the same thing have hugely disparate names, edit distance may perform worse than the other two similarity measures. In general, in cases where the two tables whose schemas are being matched belong to the same database, they are like to use same or similar names to address similar values and edit distance should be the choise of measure of similarity. However, in cases where the tables are coming from two different databases, there is a high possibility of the two schemas having very dissimilar names, in which case edit distance would be a poor choice of similarity measure. 

In conclusion, while the methods proposed do not outperform the ones currently in use, given the small amount of data needed, they can be put to practical use, especially in systems where similarity between attribute names can be exploited. 

\section{Future Work}
At present, the one to one mappings performed are relatively simple and do not allow complex mappings to take place. We can extend this work by taking inspiration from \cite{ref5} and creating searchers that can take care of these complex mappings as well. The process of dictionary creation can also be improved so that the process can become semi-automated rather than being completely manual. One possiblity could be to scrape the internet to find commonly occurring one-to-many mappings and already have them included in the dictionary. Another method could be to exploit the natural language processing tasks like n-gram creation to create a space of all possible n-grams that could be included in the on-to-many mapping dictionary.

\ifCLASSOPTIONcaptionsoff
  \newpage
\fi



%
\bibliographystyle{plain}
\bibliography{egbib.bib}

\end{document}